\documentclass[12pt]{iopart}

\usepackage[utf8]{inputenc}
\usepackage{iopams}
 \expandafter\let\csname equation*\endcsname\relax
\expandafter\let\csname endequation*\endcsname\relax
\usepackage{amsfonts,amsmath,amssymb,graphicx,epstopdf,verbatim}
\usepackage{iopams}
\usepackage{placeins}
\usepackage{rotating}
\usepackage{xcolor}
\usepackage{braket}
\usepackage{enumitem}
\usepackage[colorlinks]{hyperref}

\definecolor{darkred}{rgb}{0.5,0,0}
\definecolor{darkgreen}{rgb}{0,0.5,0}
\definecolor{darkblue}{rgb}{0,0,0.5}
\hypersetup{colorlinks, linkcolor=darkred, filecolor=darkgreen, urlcolor=darkblue, citecolor=darkgreen}

\UseRawInputEncoding

\newcommand{\mb}{\mathbb}

\begin{document}

\title{The Ising critical quantum Otto engine}

\author{Giulia Piccitto}
\address{Dipartimento di Fisica dell'Universit\`a di Pisa and INFN, 
  Largo Pontecorvo 3, I-56127 Pisa, Italy}

\author{Michele Campisi}
\address{NEST, Istituto Nanoscienze-CNR and Scuola Normale Superiore,
  I-56127 Pisa, Italy}

\author{Davide Rossini}
\address{Dipartimento di Fisica dell'Universit\`a di Pisa and INFN, 
  Largo Pontecorvo 3, I-56127 Pisa, Italy}

\begin{abstract}
  We study a four-stroke Otto engine whose working fluid is a quantum Ising chain.
  The thermodynamic cycle consists in sweeps of the transverse magnetic field occurring in thermal
  isolation, alternated by thermalisation strokes with reservoirs at different temperatures.
  The system-environment coupling is modelled in a thermodynamically consistent way by means of a
  nonlocal Lindblad master equation.
  We show that the engine may operate in four different operation modes, depending on
  the various parameters, in particular it can act as a heat engine and as a refrigerator. 
  We detect an enhancement of the thermodynamic performance as the critical point is crossed,
  and investigate it in detail.
\end{abstract}

\maketitle

\section{Introduction}
\label{sec:intro}

In everyday life we are continuously in contact with engines. These are systems of classical interacting particles that, under suitable thermodynamic transformations, can both convert heat in work and viceversa. 
Heat engines, machines that convert heat into useful work (and viceversa), are ubiquitous in our everyday life.
Their working mechanism can typically be well described within the framework of classical physics.
However, around the mid of last century, it has become clear that heat engines may as well be based
on a genuinely quantum-mechanical working substance, the prime example being the maser~\cite{Scovil59PRL2}.
Later on, in the early 1980s, further pioneering proposals of quantum engines 
have been put forward~\cite{Alicki1979, Kosloff1984}, thus initiating a prolific research
field~\cite{Quan2007, Scully2011, linden2010, allahverdyan2008work, Kosloff-rev2014, Friedenberger2017, Fialko2012, Rossangel2014, Niedenzu2016, Uzdin2016, Hovhannisyan2013, giorgi2015}.
Recently, a series of works in this streamline have focused on the possibility to have few-body
(e.g., one- or two-qubit) quantum engines undergoing a Carnot
cycle~\cite{Bender2000, bender2002entropy, Kurizki2013, Abiuso2020} or an Otto
cycle~\cite{Henrich2007, Uzdin2014, Leggio2016, Kosloff-rev2017, Mehta2017, Alvarado2017, Solfanelli2020, DelGrosso2022, Abah2012}.
Some of them have been realised in the lab, with a number of distinct experimental platforms,
including trapped ions~\cite{SingleAtomExp2016, TrappedIons2019, vonLindenfels2019},
NMR~\cite{Peterson2019}, NV centers~\cite{Klatzow2019}, cavity optomechanics~\cite{Sheng2021},
ultracold atoms~\cite{Bouton2021}, as well as the superconducting qubits of a quantum
processor~\cite{Solfanelli21PRXQ2,Solfanelli22arXiv220113319}---see also the recent review in Ref.~\cite{Myers22AQS4}.

\begin{figure}[!t]
  \centering
  \includegraphics[width=0.7\columnwidth]{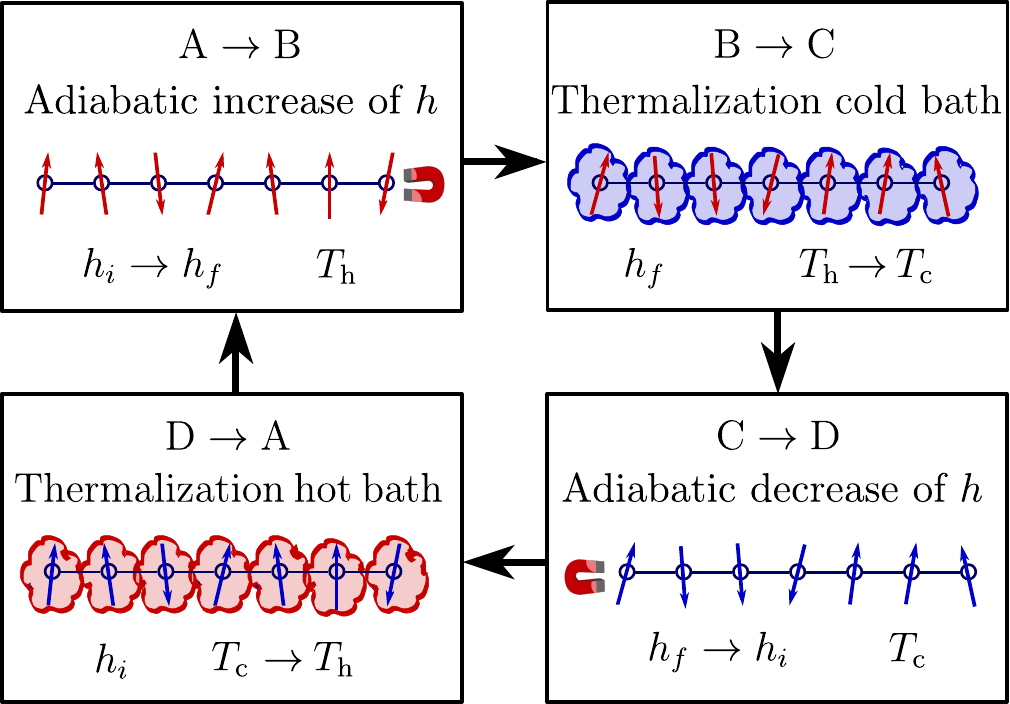}
  \caption{Sketch of the Ising quantum Otto engine (see Sec.~\ref{sec:oc} for details).
    The Ising chain is initialized in a thermal state of temperature $T = T_{\rm h}$ and subject to 
    the transverse field $h = h_i$. The thermodynamic cycle consists in the following steps:
    (${\rm A} \to {\rm B}$) the system is adiabatically driven towards $h_f > h_i$;
    (${\rm B} \to {\rm C}$) each spin is then coupled to a single bath at temperature $T_{\rm c}$
    and let thermalise;
    (${\rm C} \to {\rm D}$) the system, now being in the thermal state at $T = T_{\rm c}$
    and transverse field $h = h_f$, is adiabatically driven towards $h_i < h_f$;
    (${\rm D} \to {\rm A}$) each spin is finally coupled to another single bath at temperature
    $T_{\rm h}$ until thermalisation occurs, thus closing the cycle.}
  \label{Fig:otto_ising}
\end{figure}

However, the role that many-body interactions play in the thermodynamic performance of quantum engines
is still not fully understood: identifying whether quantum engines featuring many-body interaction may
outperform classical ones, and under which condition that happens, is currently in the limelight of intense
quantum thermodynamics debate~\cite{niedenzu2018cooperative, jaramillo2016quantum, Campaioli2017}.
One of the main obstacles is represented by the difficulties that one typically encounters
in solving the many-body system
dynamics~\cite{Mukherjee2021, Sun2017, Chen2018, Halpern2019, Chen2019, delcampo2020, Iemini2021}.
In this paper we contribute this flourishing field of research by presenting a detailed study
of the Ising quantum Otto engine (see Fig.~\ref{Fig:otto_ising}), particularly focusing on the role
that quantum criticality may have on the engine thermodynamic performance. 

It has been suggested that the divergence of equilibrium fluctuations (hence of linear response coefficients,
as per the fluctuation-dissipation theorem~\cite{Kubo66RPP29}) in proximity of a classical phase transition
could result in an enhancement of the performances of a heat engine~\cite{CampisiFazio2016}. 
An interesting question that has been only partially addressed is whether such an enhancement may be caused
as well by a quantum phase transition. A demonstration of this quantum critical enhancement has been given
for an engine made of a Tonk-Girardeau gas at the verge of the pinning-transition~\cite{Fogarty2020}
and for a Dicke quantum engine in correspondence of the superradiant critical point~\cite{Fusco2016}.
Here we provide further evidence that the presence of a quantum phase transition can lead to genuine
quantum many-body enhancement by studying an heat engine with a quantum Ising chain in transverse field
(a prototypical model exhibiting a quantum critical point~\cite{Sachdev2011}).
Specifically we observe a super-extensive scaling of the device performance, defined as~\cite{CampisiFazio2016}
\begin{equation}
  \Pi = W/ \delta \eta,
  \label{Eq:pi}
\end{equation}
where $\delta \eta = \eta_C - \eta$ denotes the difference between the engine efficiency and that
of an ideal Carnot engine, and $W$ is the work output. In absence of cooperative enhancement, the performance
scales linearly in the system size $N$, as is for $N$ identical engines working in parallel.
Superextensive scaling signals a genuine cooperative boost~\cite{CampisiFazio2016}.

We also found that the Ising quantum Otto engine can operate both as a heat engine (i.e., by absorbing
heat from a cold reservoir and partially converting it into mechanical work) and as a refrigerator
(i.e., by using the work performed on it to transfer heat from a cold reservoir to an hot one)
depending on the parameters defining the thermodynamic cycle. Even though the operation
regimes are parameter dependent, it is always possible to find a range of parameters that realises
any of them with good stability. 

The paper is organized as follows. In Sec.~\ref{sec:oc} we define our microscopic model of four-stroke
Otto engine, whose working fluid is constituted by a quantum Ising spin chain
coupled to two thermal reservoirs. After analyzing the allowed working modes of such engine
(Sec.~\ref{sec:regimes}), we concentrate on possible role of quantum criticality on its performances
(Sec.~\ref{sec:Criticality}), with emphasis on the heat engine and the refrigerator mode.
In Sec.~\ref{sec:adiabatic_issue} we comment on the role of non-perfectly relaxation in the thermalisation strokes of the cycle.
Our conclusions are drawn in Sec.~\ref{sec:Concl}.
The Appendices provide technical details on the diagonalization of the quantum Ising chain
(\ref{App:ising}) and the modelization of external thermal baths in the adopted Born-Markov
and secular approximations (\ref{sec:therm}).

\section{The quantum Otto cycle}
\label{sec:oc}

The engine discussed in this paper operates between two temperatures $T_{\rm c}$
and $T_{\rm h}$ (with $T_{\rm c} < T_{\rm h}$), and features a quantum Ising chain as the working substance.
Its Hamiltonian reads:
\begin{equation}
  {\hat H_{\rm sys}}(t) = -J\sum_{j=1}^{N-1} \hat \sigma_j^x \hat \sigma_{j+1}^x - h(t) \sum_{j=1}^N \hat \sigma_j^z,
  \label{Eq:ising}
\end{equation}
with $\hat \sigma_j^\alpha$ being the spin-$1/2$ Pauli operators acting on the $j$th site ($\alpha = x,y,z$). It
describes a one-dimensional Ising system of $N$ quantum spins interacting with a ferromagnetic coupling
strength $J>0$, in the presence of a possibly time-dependent transverse magnetic field $h(t)$.
Hereafter we set $J=1$ as the energy scale and work in units of $\hbar = k_B = 1$.

After initializing the chain in the thermal state at temperature $T_{\rm h}$
and transverse magnetic field $h=h_i$, the following four strokes are implemented sequentially for each cycle,
as sketched in Fig.~\ref{Fig:otto_ising} (we set $h_i < h_f$, since the energy of the Ising system~\eqref{Eq:ising}
at sufficiently low temperatures decreases with the transverse field --- cf.~\ref{App:ising}):
\begin{enumerate}
\item {\bf ${\rm A} \to {\rm B}$: Adiabatic increase of the field.}
  The transverse field is quenched linearly in time
  from $h_i$ to $h_f$, as in Eq.~\eref{eq:linquench}, 
  while keeping the working substance decoupled from the baths;
\item {\bf ${\rm B} \to {\rm C}$: Thermalisation with the cold bath.}
  The Hamiltonian $\hat H_{\rm sys}(t_f)$ is kept fixed, while the coupling with
  the hot bath is turned on, until the working substance is described by the thermal state
  of $\hat H_{\rm sys}(t_f)$ at temperature $T_{\rm c}$;
\item {\bf ${\rm C} \to {\rm D}$: Adiabatic decrease of the field.}
  The transverse field is linearly quenched back from $h_f$ to $h_i$,
  with the same velocity as in stroke 1, while keeping the working substance decoupled from the baths;
\item {\bf ${\rm D} \to {\rm A}$: Thermalisation with the hot bath.}
  The Hamiltonian $\hat H_{\rm sys}(t_i)$ is kept fixed, while the coupling with
  the cold bath is turned on, until the working substance comes back to the
  initial thermal state of $\hat H_{\rm sys}(t_i)$ at temperature $T_{\rm h}$.
\end{enumerate}

The two adiabatic strokes acting on the quantum Ising chain
are implemented by linearly varying the transverse field $h$ 
from an initial value $h_i$ to a final value $h_f$, with a velocity $v$, as
\begin{equation}
  h(t) = h_i + v t , \qquad t \in [0, (h_f-h_i) / v] ,
  \label{eq:linquench}
\end{equation}
and vice-versa. 
%
The work performed by the system during such sweep is 
\begin{equation}
  W_{i \to f} = -\int_{t_i}^{t_f} dt' \ \frac{d{\braket{\hat H(t')}_{\rho(t')}}}{{dt'}} 
  = \braket{\hat H(t_i)}_{\rho(t_i)} - \braket{\hat H(t_f)}_{\rho(t_f)},
  \label{eq:WorkAd}
\end{equation} 
where $\braket{\, \cdot \,}_{\rho(t)}$ denotes the expectation value over the state $\rho(t)$ of the system
at time $t$ (not necessarily pure). In going from the first to the second line, we used the fact that
the system evolves unitarily during the adiabatic strokes.

In the following we shall use the symbol $Q_{\rm c(h)}$ to denote the heat exchanged during the 
thermalisation with the cold (hot) bath, with the convention that the system is absorbing heat from the 
reservoir if $Q_{\rm c(h)} > 0$. 
If $\rho_\alpha$ denotes the system's density matrix at point $ \alpha = A, B, C, D$, then:
\begin{subequations}
\begin{eqnarray}
  Q_{\rm c} & = & \braket{\hat H(t_f)}_{\rho_C} - \braket{\hat H(t_f)}_{\rho_B},  \label{eq:q1} \\
  Q_{\rm h} & = & \braket{\hat H(t_i)}_{\rho_A} - \braket{\hat H(t_i)}_{\rho_D}.  \label{eq:q2}
\end{eqnarray}
\end{subequations}
Note that no heat is exchanged during the adiabatic strokes and no work is performed during the
thermalisation strokes. Thus, the knowledge of the system's internal energy at each point, $A,B,C,D$,
of the cycle (evaluated by Jordan-Wigner transforming Eq.~\eref{Eq:ising} into a fermionic model and using the properties of fermionic Gaussian states, as detailed in~\ref{App:ising}) suffice for its thermodynamic characterisation.

\section{Working regimes of the Otto engine}
\label{sec:regimes}

We first analyze which kind of operation modes the Otto cycle proposed above can realise.
By combining the Clausius inequality with the first law of thermodynamics, it can be shown
that only the following four working regimes are allowed~\cite{Solfanelli2020}:
\begin{itemize}
\item Refrigerator (R): the engine absorbs energy and transfers heat from the cold reservoir to the hot one,
  i.e., $Q_{\rm c} > 0$, $Q_{\rm h} < 0$, and $W < 0$;
\item Accelerator (A): the engine absorbs energy and transfers heat from the hot reservoir to the cold one,
  i.e., $Q_{\rm c} < 0$, $Q_{\rm h} > 0$, and $W < 0$;
\item Heat engine (E): the engine produces work by absorbing heat from the hot reservoir,
  i.e., $Q_{\rm c} < 0$, $Q_{\rm h} > 0$, and $W > 0$;
\item Heater (H): the engine absorbs energy and heats up both the hot and the cold reservoir,
  i.e., $Q_{\rm c} < 0$, $Q_{\rm h} < 0$, and $W < 0$.
\end{itemize}

\begin{figure}[!t]
  \centering
  \includegraphics[width=0.7\columnwidth]{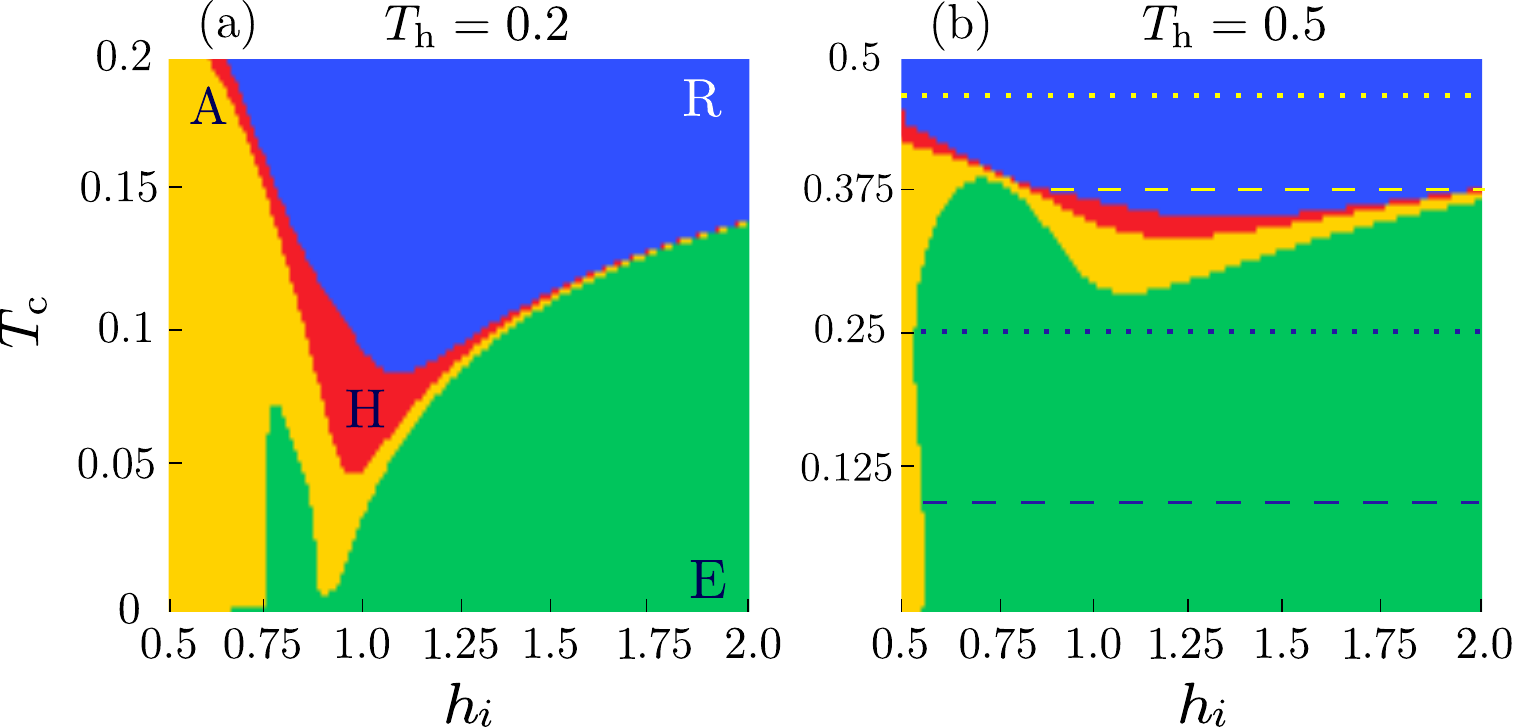}
  \caption{The various operating regimes of the quantum Otto engine, in the $h_i$--$T_{\rm c}$ plane,
    for $T_{\rm h} = 0.2$ (a) and $0.5$ (b), and fixed $\delta h = h_f-h_i = 0.5$.
    The color code stands for accelerator (yellow), heater (red), heat engine (green), and refrigerator (blue).
    Straight blue and yellow lines in panel (b) mark the values of $T_{\rm c} = 0.1, \, 0.25$
    and $0.375, \, 0.45$, respectively; these configurations are analyzed in detail below.}
  \label{Fig:regimes}
\end{figure}

Fig.~\ref{Fig:regimes} shows the Ising Otto engine ``phase'' diagrams for an Ising chain with $N = 50$ spins,
$T_{\rm h} = 0.2$ [panel (a)] and $T_{\rm h} = 0.5$ [panel (b)], for a fixed quench amplitude
$\delta h \equiv h_f - h_i = 0.5$ and fixed velocity $v = 0.005$. In what follows, unless specified,
  we assume these values of $N$, $\delta h$, and $v$ as parameters for our numerical simulations.
The plots evidence that the engine is able to operate in all the four working regimes with a good stability
(the corresponding regions are extended in the parameter space).
Contrary to the usual expectation, the heater (red area) is not the prevailing regime,
while there are wide configuration ranges allowing for the heat engine (green area) and for the refrigerator (blue area). 

\begin{figure}[!t]
  \centering
  \includegraphics[width=0.7\columnwidth]{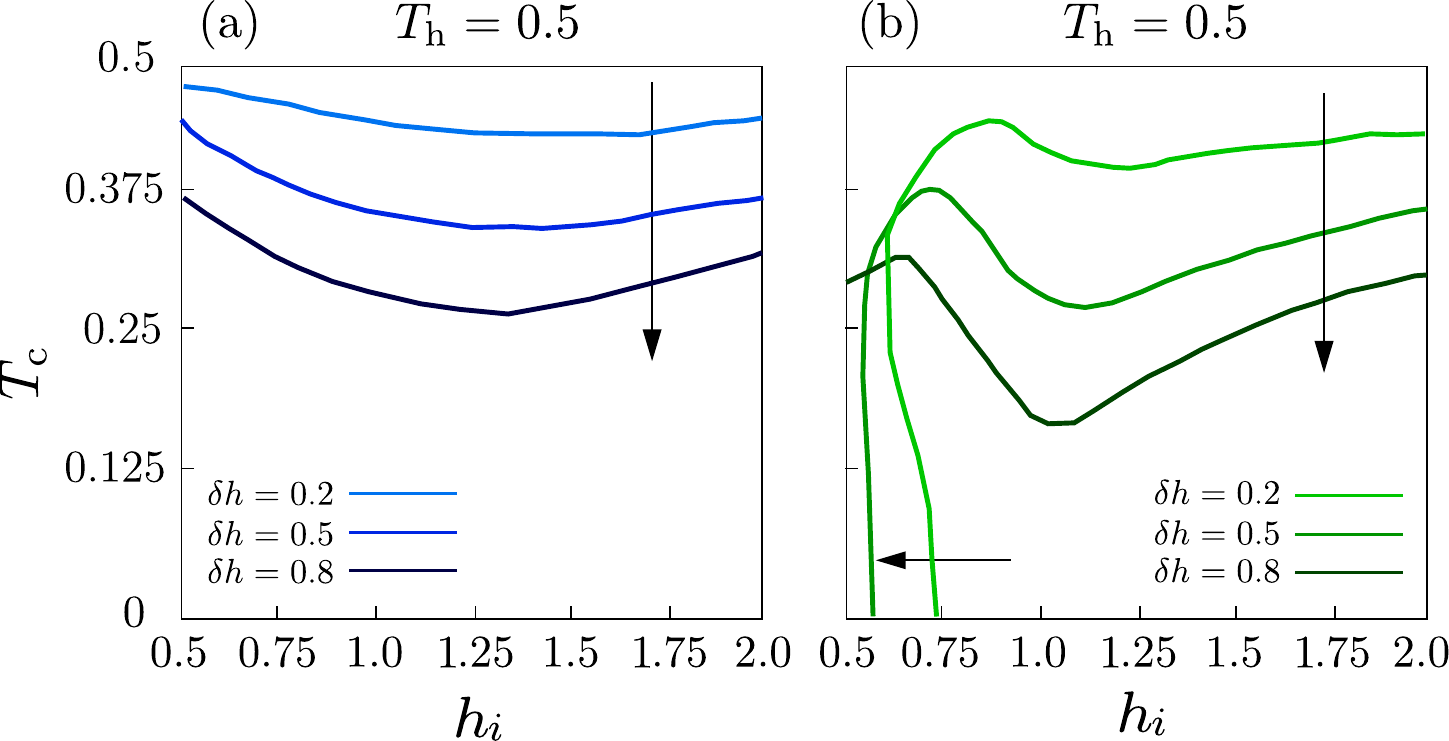}
  \caption{Boundaries of the refrigerator (a) and of the heat engine (b) regions,
    obtained for $N = 50$ and $T_{\rm h} = 0.5$, when varying $\delta h = 0.2, 0.5, 0.8$
    (from lighter to darker colors).}
  \label{Fig:boundaries}
\end{figure}

The geometry of the diagram in Fig.~\ref{Fig:regimes} depends on the system parameters,
also including the temperature $T_{\rm h}$ and the quench amplitude $\delta h$. 
The effect of the temperature is particularly noticeable in proximity of the Ising critical point
$h_{\rm crit} = 1$: the region associated with the heat engine exhibits a reentrance
that becomes more pronounced when reducing the temperature of the hottest reservoir
[compare panel (a) with panel (b)]. 
This reentrance is directly related to the presence of quantum criticality. In fact, because of the
vanishing energy gap, it is difficult to perform quenches close to the critical point
without exciting the system and dissipating energy~\cite{dorner2012}. 
On the other hand, the quench amplitude strongly affects the boundaries of the different regions. In particular, we observe that by increasing $\delta h$ the boundary of the green regions shift to the left, while the lower boundary of the refrigerator region expands with decreasing $T_{\rm c}$.
These effects are further illustrated in Fig.~\ref{Fig:boundaries}, where we show the boundary
of the refrigerator region [panel (a)] and of the heat engine [panel (b)],
when $\delta h$ is varied, for $N = 50$ and $T_{\rm h} = 0.5$. 

\section{Role of quantum criticality}
\label{sec:Criticality}

In this section we thoroughly investigate the effects of quantum criticality on the performance
of the Otto engine, explicitly focusing on the heat engine and on the refrigerator mode.

\subsection{Critical heat engine}
\label{sec:E}

\begin{figure}[!t]
  \centering
  \includegraphics[width=0.7\columnwidth]{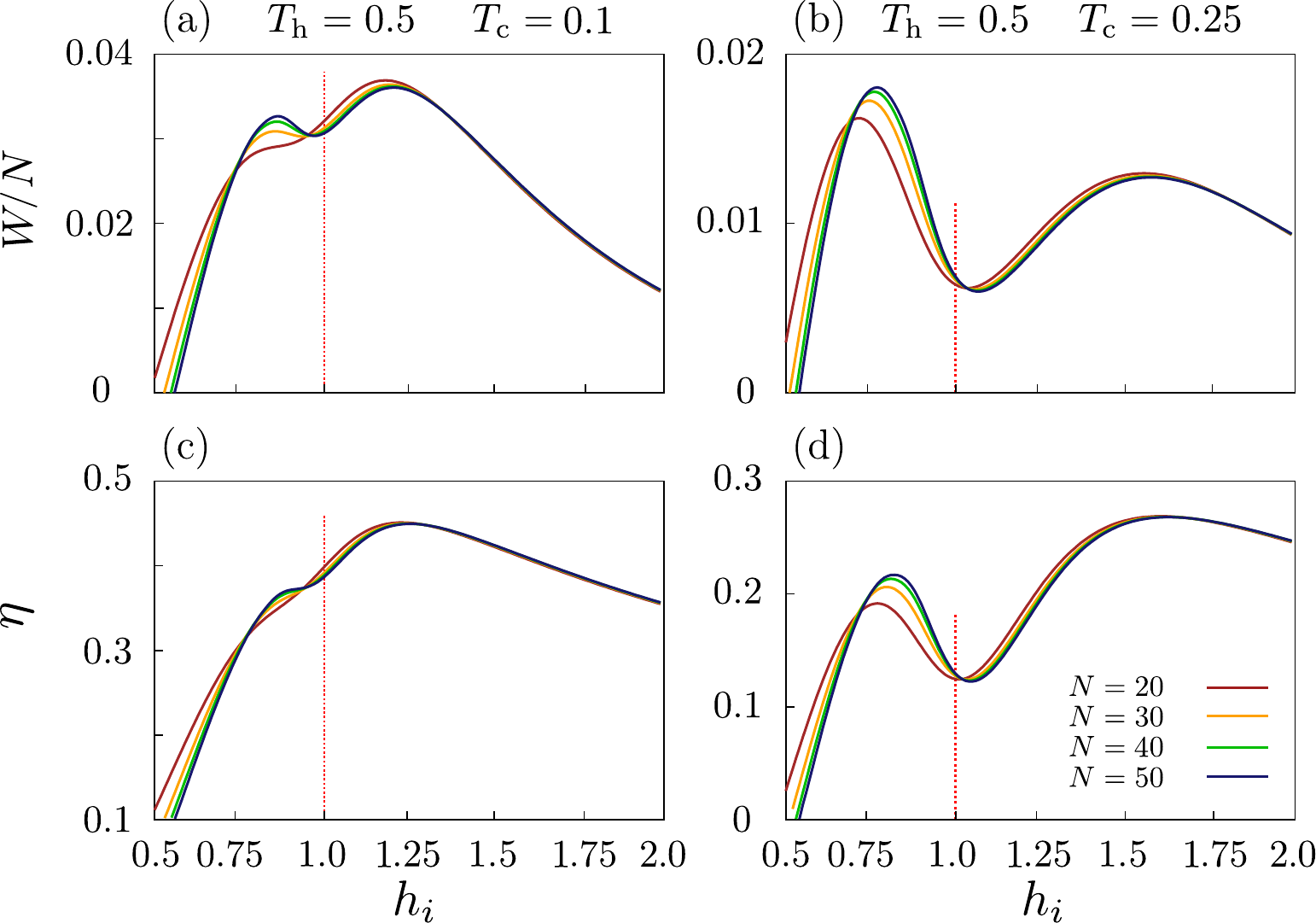}
  \caption{Top panels: Work per spin $W/N$ versus $h_i$ for different system sizes $N$, and for
    $T_{\rm c} = 0.1$ (a) and $0.25$ (b). We distinguish two peaks appearing at $h_i < h_{\rm crit}$, for quenches
    across the critical point, and at $h_i > h_{\rm crit}$, for quenches in the paramagnetic phase.
    In correspondence of the former (that becomes more resolved while increasing $T_{\rm c}$), the work exhibits
    a superextensive scaling with $N$, hinting at a possible role of quantum criticality
    in the system. Bottom panels: same as in the top panels, but for the efficiency of the heat engine $\eta$,
    which shows the same qualitative structure observed in the top panels,
    including the dependence on $N$ in correspondence of the critical peak.
    Red dotted lines mark the point $h_{\rm crit} = 1$.
    In these figures we fix $T_{\rm h} = 0.5$ and $\delta h=0.5$.}
  \label{Fig:WvsQ}
\end{figure}

The two quantities characterising the performance of a heat engine are the work output $W$
and the efficiency $\eta$, the latter being defined as the ratio of the work output
over the heat extracted,
\begin{equation}
  \eta = W/Q_{\rm h}.
\end{equation}
Here we are most interested in finding and analyzing the parameters range that maximizes
both $W$ and $\eta$. Note that the second law of  thermodynamics forces the latter to be bounded
by the Carnot efficiency: $\eta \le \eta_{C} = 1 - T_{\rm c}/T_{\rm h}$. 

In Fig.~\ref{Fig:WvsQ} we show, for various system sizes, the work per spin (top panels) and the efficiency
(bottom panels) versus the initial transverse field, for fixed $T_{\rm h} = 0.5$ and $T_{\rm c} = 0.1, \, 0.25$
[dashed and dotted blue line in Fig.~\ref{Fig:regimes}(b), respectively]. The first emerging feature
is that both the work and the efficiency have a double-peak structure that becomes more resolved
while increasing $T_{\rm c}$.  One of the two peaks appears at $h_i < 1$, in correspondence to quenches
across the critical point ($h_{\rm crit} = 1$),
while the other one at $h_i > 1$, when considering quenches in the paramagnetic phase. 
For convenience, we refer to them as the {\em critical} and the {\em paramagnetic} peak, respectively.
The paramagnetic peak is substantially independent of the system size, meaning that a $N$-body engine
behaves as $N$ one-body engines.
In contrast, the critical peak displays a non trivial dependence on $N$, suggesting the possibility
that quantum criticality may enhance the cooperative effects.

As discussed in Ref.~\cite{CampisiFazio2016}, when considering finite-temperature systems close to criticality,
it is reasonable to expect an improvement of the heat extraction in correspondence of the critical point. 
In fact, because of the divergence of the specific heat, a critical system can exchange a large amount
of heat even in presence of a small gradient of temperature. 
A similar argument can hold in our magnetic system. 
When approaching the quantum critical point, the magnetic susceptibility (defined as the derivative
of the magnetization with respect to the associated field) diverges, meaning that
the magnetization becomes very sensitive also to small changes of the field. 
The work associated to a change in the magnetic field is proportional to the
magnetization~\cite{dorner2012, fusco2014}, therefore we expect the work extraction to be improved
in correspondence of the quantum critical point.

\begin{figure}[!t]
  \centering
  \includegraphics[width=0.9\columnwidth]{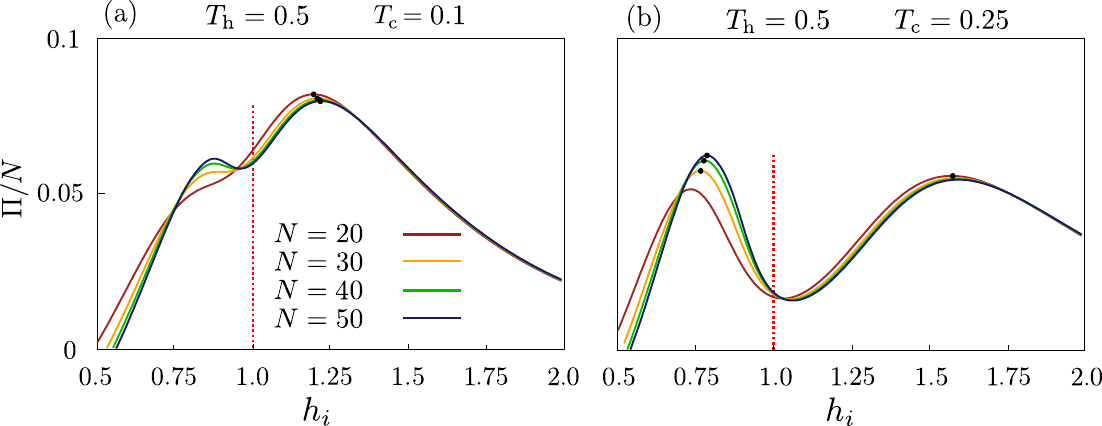}
  \caption{Behavior of $\Pi/N$ versus $h_i$, for different system sizes $N$, and for $T_{\rm c} = 0.1$ (a)
    and $0.25$ (b). The dots mark the maxima of $\Pi/N$ for each value of $N$.
    In both panels we recognize the same double-peak structure previously discussed in Fig.~\ref{Fig:WvsQ}.
    Red dotted lines mark $h_{\rm crit} = 1$.}
  \label{Fig:pi}
\end{figure}

The critical improvement of the engine performance is evidenced in Fig.~\ref{Fig:pi},
showing the behavior of the thermodynamic performance $\Pi$, as defined in Eq.~\eref{Eq:pi},
as a function of $h_i$ and for the same parameters of Fig.~\ref{Fig:WvsQ}.
The dots indicate the points at which the absolute maxima of $\Pi$ occur, for the various system sizes. 
We observe that, when considering small $T_{\rm c}$ [cf.~$T_{\rm c} = 0.1$ in Fig.~\ref{Fig:pi}(a)],
the maxima are all located in correspondence of the paramagnetic peak. 
In contrast, when increasing $T_{\rm c}$ [cf.~$T_{\rm c} = 0.25$ in Fig.~\ref{Fig:pi}(b)], the absolute maxima
of $\Pi$ distribute partly on the paramagnetic peak (for $N=20$)
and partly on the critical peak (for larger values of $N$), depending on the system size.

This means that, in the presence of a large gradient of the reservoir temperatures, to maximize the performances
it is convenient to perform quenches inside the paramagnetic phase. 
When increasing the temperature of the cold bath, instead, the work production far from the critical point
suffers from the reduced heat absorption due to the reduced reservoir gradient. 
In this case, the maxima of $\Pi$ move on the critical peak (intuitively the work extraction loses efficiency
far from the critical point, while it improves in proximity of $h_i = h_{\rm crit}$). 
However, it is not easy to determine {\it a priori} whether the absolute maximum of $\Pi$
is in correspondence of the critical or of the paramagnetic peak. 
Moreover, since criticality emerges with increasing the system size, it could be necessary to consider
relatively large values of $N$ to exploit the effects of critical enhancement.
Interestingly, we notice that, while the height of the paramagnetic peak in $\Pi/N$ evidences an ordinary
linear scaling $\Pi \sim N$, the height of the critical peak appears to scale more than linearly.
The details of this scaling behavior will be discussed more thoroughly in Sec.~\ref{subsec:qec}.

\subsection{Critical refrigerator}
\label{sec:R}

The above analysis can be naturally extended to the parameter range where the engine operates
in the refrigerator mode. In this case, the relevant quantities to be considered are the heat $Q_{\rm c}$
extracted from the cold reservoir, and the coefficient of performance (COP)
\begin{equation}
  \eta^R = Q_{\rm c}/W.
\end{equation}
As for the heat engine, due to the second law of thermodynamics, this ratio cannot be larger
than the Carnot COP: $\eta^R \le \eta^R_{C} = T_{\rm c}/(T_{\rm h} -T_{\rm c})$. 

\begin{figure}[!t]
  \centering
  \includegraphics[width=0.7\columnwidth]{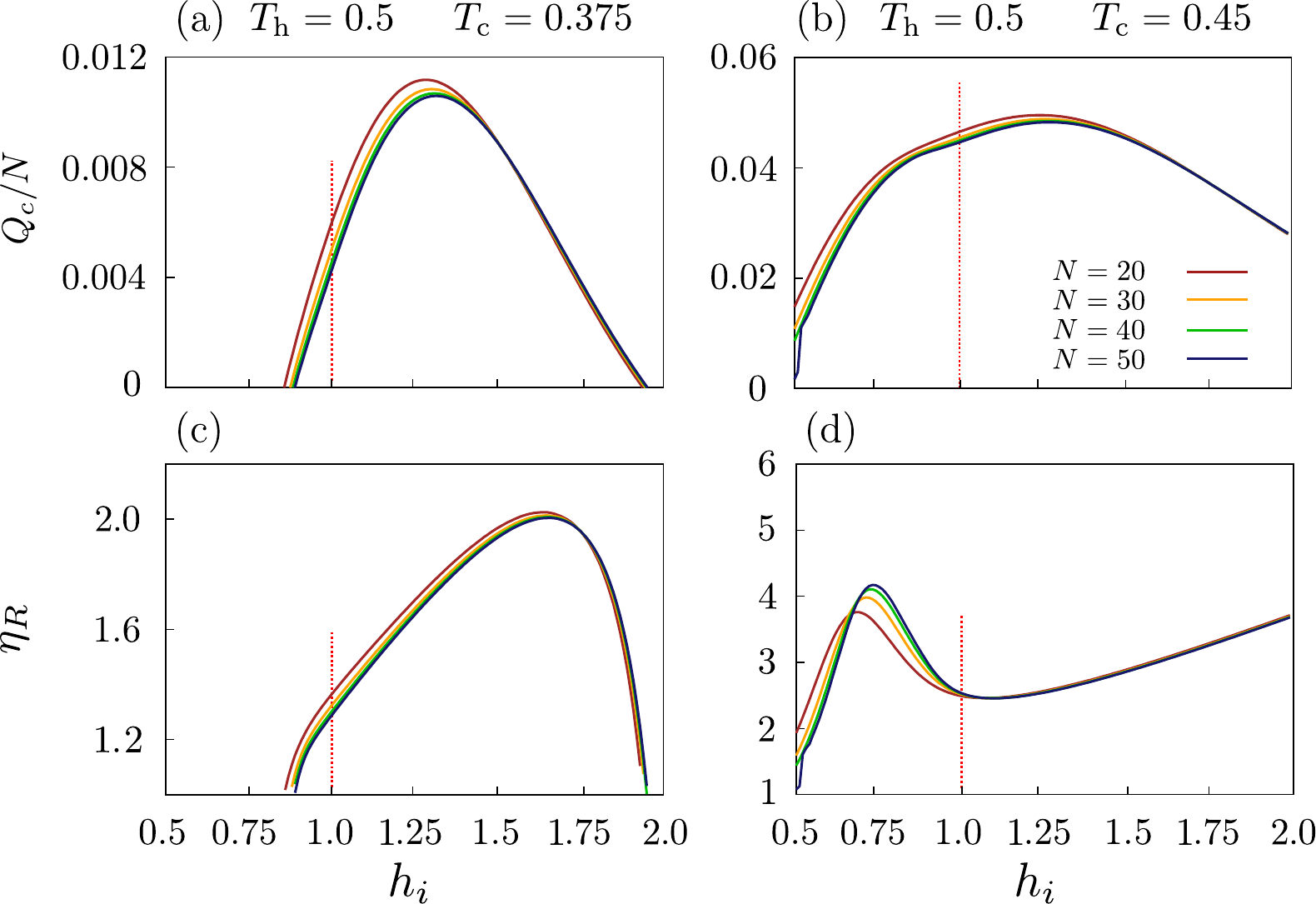}
  \caption{Top panels: Heat extracted from the cold reservoir per spin $Q_{\rm c}/N$ versus $h_i$,
    for different system sizes $N$, and for $T_{\rm c} = 0.375$ (a) and $0.45$ (b).
    The double-peak structure observed in Fig.~\ref{Fig:WvsQ} for the heat engine is now almost lost.
    Bottom panels: same as in the top panels, but for the COP of the refrigerator.
    Differently from the heat exchanged, at large $T_{\rm c}$, the efficiency shows an
    enhancement of the performance in correspondence of quenches across the critical point.
    Red dotted lines mark $h_{\rm crit} = 1$.}
  \label{Fig:QvsQ}
\end{figure}

In Fig.~\ref{Fig:QvsQ} we show, for various system sizes, the heat $Q_{\rm c}$ per spin (top panels)
and the COP (bottom panels) versus the initial transverse field $h_i$, for fixed $T_{\rm h} = 0.5$
and $T_{\rm c} = 0.375, \, 0.45$ [dashed and dotted yellow line in Fig.~\ref{Fig:regimes}(b), respectively].
We notice that the heat extracted from the cold reservoir is (almost everywhere) a convex function,
with a weak dependence on the system size $N$.
Despite this, we observe some evidence of criticality when looking at the COP for large values of $T_{\rm c}$. 
In particular, in Fig.~\ref{Fig:QvsQ}(d) we observe a well resolved peak at $h_i < 1$, displaying
a dependence on $N$ that, in analogy with the discussion of Sec.~\ref{sec:E}, reflects
the presence of quantum criticality.

\begin{figure}[!t]
  \centering
  \includegraphics[width=0.9\columnwidth]{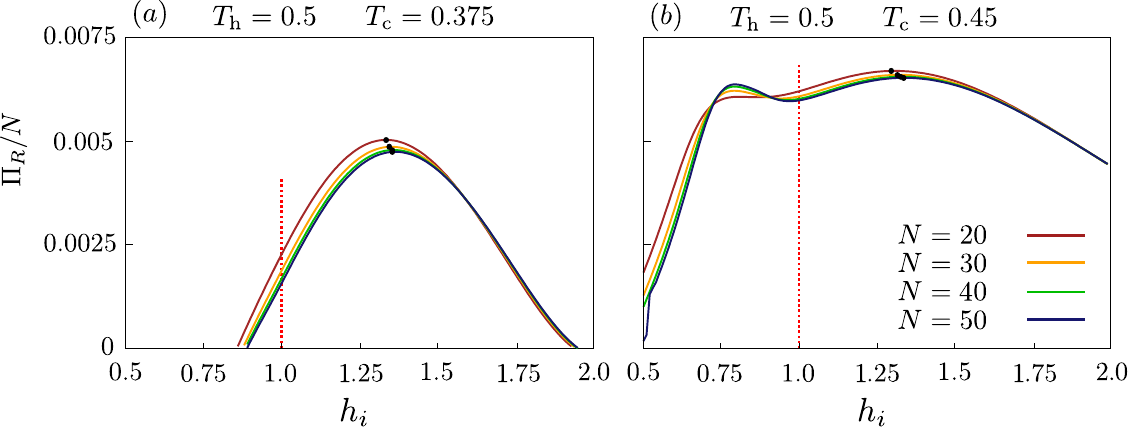}
  \caption{Behavior of $\Pi_R/N$ versus $h_i$, for different system sizes, for $T_{\rm c} = 0.375$ (a)
    and $0.45$ (b). Colored squares mark the maxima of $\Pi_R$ for each value of $N$.
    As expected from the results shown in Fig.~\ref{Fig:QvsQ}, 
    an enhancement appears in the critical peak only for $T_{\rm c} = 0.45$ and for large sizes $N$.
    Red dotted lines mark $h_{\rm crit} = 1$.}
  \label{Fig:cop}
\end{figure}

In analogy with Eq.~\eref{Eq:pi}, we introduce the quantity
\begin{equation}
  \Pi_R = Q_{\rm c} / \delta \eta^R,
\end{equation}
as a quantifier of the refrigerator performance. Here $\delta \eta^R = \eta^R_C - \eta^R$ is the difference
between the Carnot COP and the engine one. Figure~\ref{Fig:cop} shows $ \Pi_R/N$ for the same parameters
as in Fig.~\ref{Fig:QvsQ}. Colored squares mark the absolute maxima of $\Pi_R/N$. 
We notice that, differently to what discussed in Sec.~\ref{sec:E}, for the parameters we considered,
the best performances are always achieved inside the paramagnetic phase.

\subsection{Enhancement of the critical peak with the system size}
\label{subsec:qec}

\begin{figure}[!t]
  \centering
  \includegraphics[width=0.8\columnwidth]{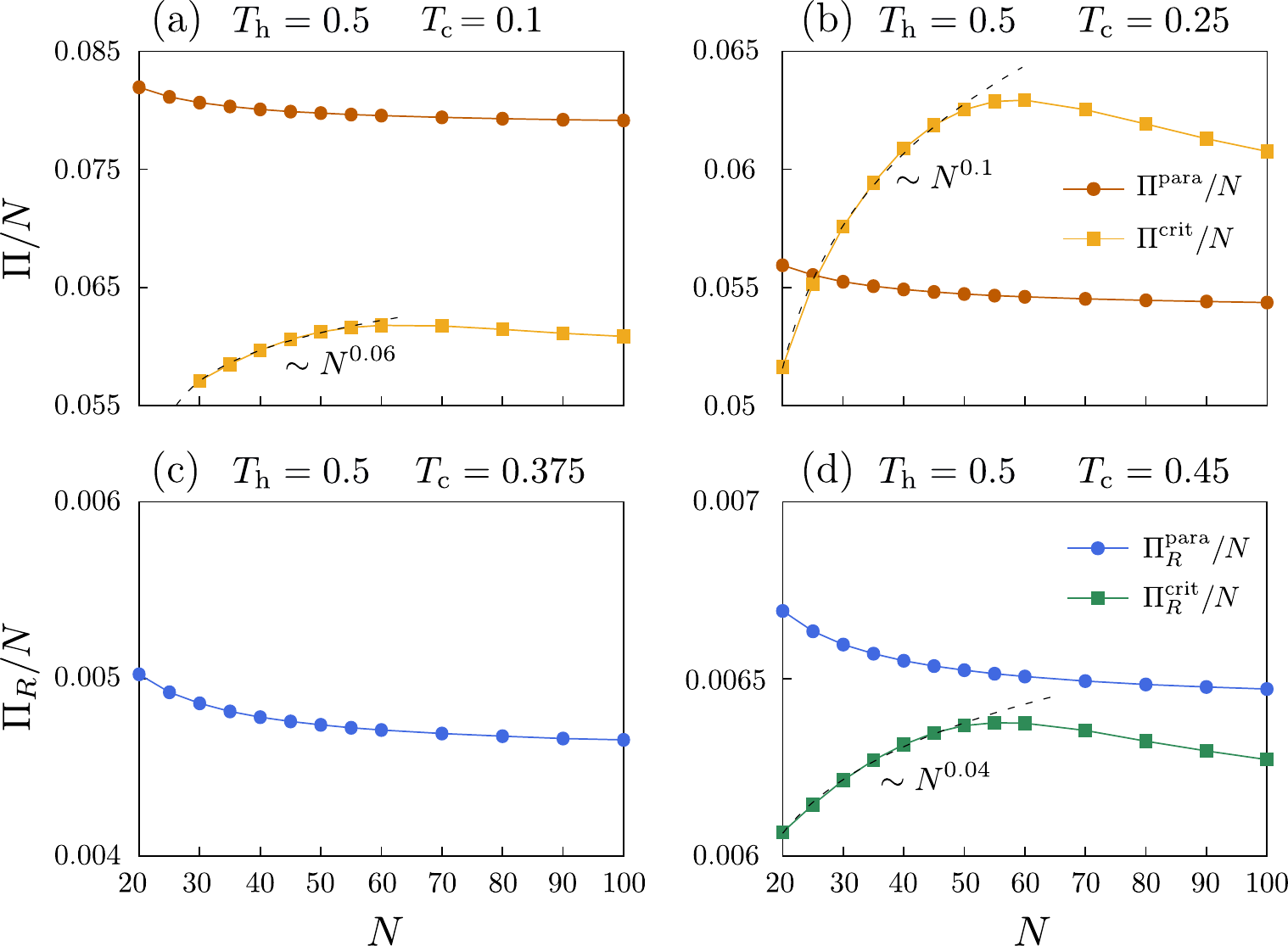}
  \caption{Top panels: behavior of the maxima of the critical (yellow squares) and of the paramagnetic
    (orange dots) peak of $\Pi/N$ of Fig.~\ref{Fig:pi} versus $N$, for $T_{\rm c} = 0.1$ (a) and $0.25$ (b).
    Bottom panels: behavior of the maxima of the critical (green squares) and of the paramagnetic
    (blue dots) peak of $\Pi_R/N$ of Fig.~\ref{Fig:cop} versus $N$, for $T_{\rm c} = 0.375$ (c) and $0.45$ (d),
    The corresponding values of $h_i$, for each data point, are read from the position
    of the peaks in Figs.~\ref{Fig:pi} and~\ref{Fig:cop}.
    Black dashed lines are power-law fits to the scaling of the critical maxima as
    $\Pi^{\rm crit}_{(R)}/N \sim N^\alpha$, with $\alpha > 0$ depending on the system parameters. 
    Notice the absence of the critical peak for $T_{\rm c} = 0.375$.} 
  \label{Fig:pi_vs_N}
\end{figure}

The presence of an enhancement of the critical peak clearly emerges when looking at the behavior of the performance
as a function of the system size. 
As already mentioned above, the height of the paramagnetic peak in $\Pi/N$ evidences an ordinary
linear scaling $\Pi \sim N$, while the critical peak appears to scale more than linearly. 
In Fig.~\ref{Fig:pi_vs_N} we show the maxima of both peaks of $\Pi/N$ for $T_{\rm c} = 0.05, 0.25$
(top panels), and of $\Pi_R/N$ for $T_{\rm c} = 0.375, 0.45$ (bottom panels), as functions of $N$. 
As expected, the former smoothly decreases while increasing $N$, to eventually settle
to a constant value. In contrast, the maxima of the critical peak follow a more than linear scaling
that is well fitted (black dashed lines) by the power law
\begin{equation}
	\Pi^{\rm crit}_{(R)}/N \sim N^\alpha, \qquad (\alpha > 0) .
\end{equation}
This scaling is not universal and the value of $\alpha$ depends on the system parameters.
In particular we observed that, for fixed $T_{\rm c}/T_{\rm h}$, the exponent $\alpha$ increases when cooling down the system (i.e., when the weight of the ground state increases), corroborating the hypothesis of an enhancement of the performance due to criticality. 

We point out that the above trend is suppressed when moving toward larger system sizes. 
In fact, the enhancement of the critical peak is a crossover effect related with
the closure of the gap $\Delta$ in the spectrum, in correspondence to the critical point.
According to the Landau-Zener mechanism~\cite{Landau1932, Zener1932}, we predict it to appear when
the quench velocity is of the order $v \sim \Delta^2$: otherwise, if changes are too fast
(sudden quench limit) or too slow (quantum adiabatic limit) compared with the gap of the system,
the dynamics becomes insensitive to the presence of the closure of the gap and thus of criticality.

\begin{figure}[!t]
  \centering
  \includegraphics[width=0.8\columnwidth]{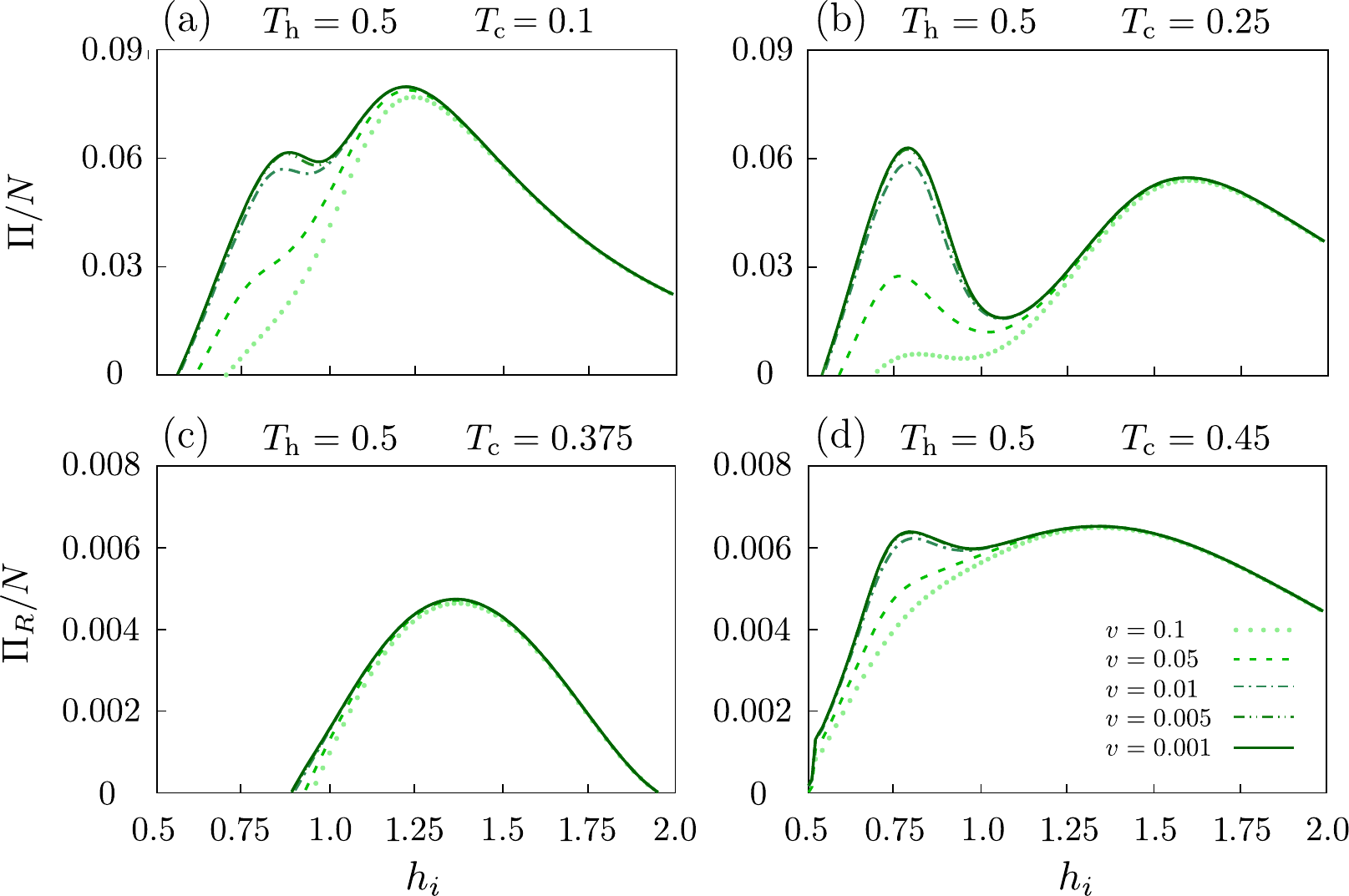}
  \caption{$\Pi/N$ (top panels) and $\Pi_R/N$ (bottom panels) versus $h_i$, for $T_{\rm c} = 0.1$ (a),
    $0.25$ (b), $0.375$ (c), and $0.45$ (d), for $N = 50$ and different quench velocities $v$ (see legend). 
    The critical peak grows when reducing the quench velocity, to eventually saturate at a maximum value.}
  \label{Fig:pi_vs_v}
\end{figure}

This is shown in Fig.~\ref{Fig:pi_vs_v}, which displays the performances $\Pi$ (cf. Fig.~\ref{Fig:pi})
and $\Pi_R$ (cf. Fig.~\ref{Fig:cop}) at $N = 50$ ($\Delta^2 \sim 4 \times 10^{-3}$),
for different quench velocities (down to $v= 10^{-3}$).
We notice that the critical peak grows when reducing the quench velocity, to eventually reach
a maximum value (thus we still do not observe the quantum adiabatic limit).
Conversely, the paramagnetic peak is almost unaffected by the quench parameters.

\section{Effects of partial thermalisation}
\label{sec:adiabatic_issue}

In deriving the results of Sec.~\ref{sec:regimes} and~\ref{sec:Criticality}, we assumed the system
to be in the thermal state at the end of stroke 2 and 4.  
In this section we discuss the effects of relaxing this assumption. 
To this end, one needs to model the open quantum dynamics of the system as induced by its interaction
with the thermal baths. The total time-dependent Hamiltonian for the system coupled to the environment,
describing the Otto cycle, can be cast in the following form:
\begin{equation}
  \hat H(t) = \hat H_{\rm sys}(t) +
  \sum_{i=1,2} \big[ \hat H_{\rm env}^i + \theta_i(t) \, \hat H_{\rm int}^{i} \big] ,
\end{equation}
where $\hat H_{\rm sys}(t)$ is the Ising time-dependent Hamiltonian of Eq.~\eref{Eq:ising},
$H_{\rm env}^i$, $i=1,2$ are the Hamiltonians of the two thermal baths, and $H_{\rm int}^{i}$
describes the coupling between the system and $i$th bath.
That is, during the adiabatic strokes we have $\theta_1(t) = \theta_2(t) = 0$, so that the system
and the environment are fully disconnected; during the stroke of thermalisation with bath $1$ we have
$\theta_1(t) = 1$ and $\theta_2(t) = 0$; similarly, during the stroke of thermalisation with bath $2$ we have
$\theta_2(t) = 0$ and $\theta_1(t) = 1$.
The explicit forms of $\hat H_{\rm env}^i$ and of $\hat H_{\rm int}^i$ are provided in~\ref{sec:therm}. 

Following Ref.~\cite{DAbbruzzo2021}, we assume the thermalisation dynamics to be ruled by a non-local
Lindblad master equation (cf.~\ref{sec:therm}) that, at variance with the more common modeling
in term of a master equation with Lindblad jump operators acting locally in the physical space
of the system~\cite{ZollerRev2012}, naturally accounts for stationary thermal states, thus avoiding
possible thermodynamic inconsistencies~\cite{LevyKosloff2014}. 
This formalism provides an analytic expression for the correlation functions of the normal modes
of the system [introduced in Eq.~\eref{eq:normal}] at any moment of the relaxation process
[cf. Eq.~\eref{eq:corr_t}]. 

\begin{figure}[!t]
  \centering
  \includegraphics[width=0.7\columnwidth]{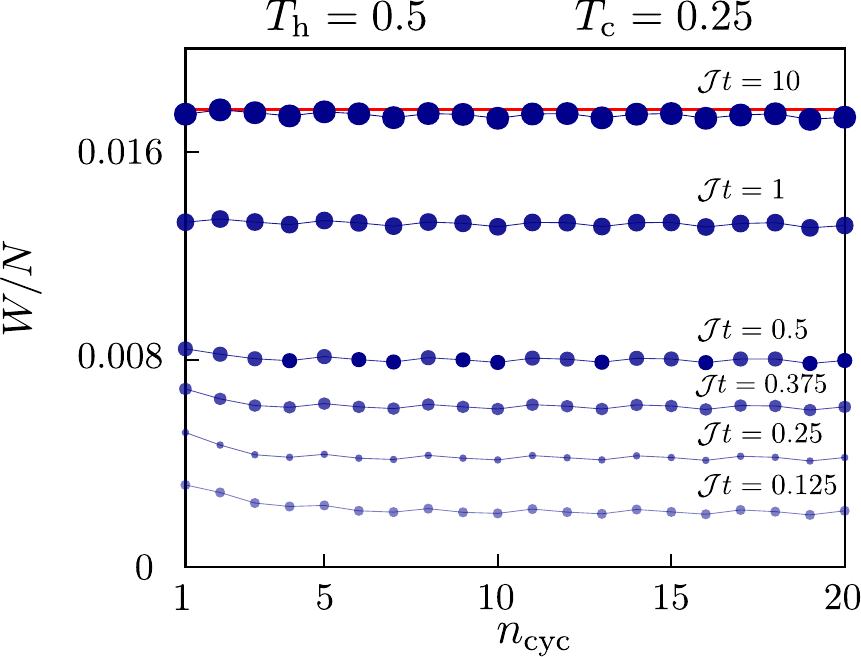}
  \caption{Work output of a non-thermalised engine of $N = 50$ spins at $T_{\rm c} = 0.25$ and $T_{\rm h} = 0.5$ in function of the number of cycles $n_{\rm cyc}$ for different thermalisation times (from lighter to darker colors). The work output, after a transient, settle on a stationary value that is bounded from the work output of the perfect thermalised engine (red line). }
  \label{Fig:w_vs_t}
\end{figure}

In particular, the state at time $t$ of a system prepared in thermal equilibrium with a reservoir
characterised by a temperature $T$ can be written as a weighted sum of two thermal states is
\begin{equation}
  \rho(t) = \rho_{T} \big( 1-e^{-2\mathcal{J}t} \big) + \rho(t=0) \, e^{-2\mathcal{J}t} ,
  \label{eq:rho_t}
\end{equation}
where $\mathcal{J}$ denotes the bath density of states and $\rho_T$ is the thermal state at temperature $T$. 
From the above expression, we find that the heat exchanged during this process reads
\begin{equation}
  Q(t) = \big[\braket{\hat H}_{\rho_T} - \braket{\hat H}_{\rho(t=0)}\big] \big(1 - e^{-2 \mathcal{J} t} \big),
\end{equation}
analogously to that of a perfect thermalised cycle reduced by a factor $1 -e^{-2\mathcal{J} t}$.

In Fig.~\ref{Fig:w_vs_t} we show the work output of a non-thermalised engine with $N = 50$ spins,
$T_{\rm h} = 0.5$, and $T_{\rm c} = 0.25$ in function of the cycle duration $n_{\rm cyc}$, for different thermalisation times.
As expected, after a transient, the work output sets into a stationary value that is bounded from above
by the work output accompanying a complete thermalisation (red line). The latter value is approached
in the long-time limit, $\mathcal J t \gg 1$.

\section{Discussion and conclusions}
\label{sec:Concl}

We presented a quantum Otto cycle with a many-body working substance made of a transverse-field
quantum Ising chain that alternately (i) evolves unitarily with a time-dependent transverse field,
or (ii) undergoes thermalisation while in contact with a thermal reservoir.
The dynamics during the thermalisation processes is ruled by a nonlocal Lindblad
master equation, which properly describes the coupling with thermal reservoirs.
We investigated the operation modes of this engine, 
finding that there are large sets of parameters for which it realises either a heat engine
or a refrigerator, depending on the reservoir temperatures and the details of the thermodynamically
adiabatic transformations.

To quantify the performances of the heat engine, we analyzed the work output, the efficiency,
and their ratio. Such indicators exhibit a double-peak structure, one in correspondence of
quenches across the quantum critical point and the other one in correspondence of quenches
inside the paramagnetic phase. 
The former becomes more resolved when reducing the temperature gradient between the two reservoirs and displays
a more-than-linear dependence on the system size, revealing the presence of quantum criticality. 
We extended the discussion to the refrigerator by considering the heat extracted from the cold reservoir,
the coefficient of performance, and their ratio. Even in this case we found evidence of quantum criticality,
although less perceptible than in the heat engine.
Finally we discussed the effects of a partial thermalisation, which provided an opportunity to understand
the physics of more realistic quantum many-body engines in contact with external baths.

Our results may serve as a useful guidance for near-term experiments with ion traps,
allowing for the realisation of interacting spin chains with $O(10^2)$ spins~\cite{Bernien2017, Zhang2017}.
While the absolute performances of the engine are, in general, maximized for quenches
inside the paramagnetic phase, the scaling of the engine with the system size $N$ can be optimised
for quenches across the critical point (although the superextensive scaling of the critical peak
saturates for large values of $N$, suggesting a finite window of $N$ to exploit
the enhancement of the performance close to criticality). 

In this work we mainly focused on the performances of the engine in terms of work done
(heat exchanged) per spin, although other aspects may be still addressed. 
Among them, we mention the problem of power-output optimisation.
This is an important point that requires a complex analysis. In fact, this optimisation should be done
both on the thermalisation protocol  and on the adiabatic one (i.e., on the quench velocity).
For example, the optimal working speed would depend on the thermalisation time that, in turn,
is affected by the microscopic details of the bath we are modeling. Moreover, it is impossible to know {\it a priori}
whether a non-perfect thermalised stroke is less performant than a thermalised one. 
Despite this, some more refined strategies to avoid the reduction of power output, due to the slowing down
of the system parameters during the adiabatic stroke, can be devised
through shortcuts to adiabaticity~\cite{STA2019, Beau2016, hartmann2020A, hartmann2020B},
variational optimisation~\cite{Cavina2018, Suri2018}, and reinforcement learning~\cite{Erdman2022}.
Beside this, an analysis of the statistical distribution of the work output may give further
useful information, even from an experimental point of view. In fact, the work output can be subject
to strong fluctuations preventing the realisation of stable engines~\cite{Holubec2017, Denzler2020, Lutz2021}. 
Even though such fluctuations should reduce while increasing the system size,
they are also expected to suffer the presence of the quantum criticality. 
Finally we mention that the method proposed here can be easily applied to situations other than
the bosonic quadratic model employed here, where different performances may emerge~\cite{Myers2020, Zheng2015}.

\ack 

We acknowledge fruitful discussions with A.~D'Abbruzzo. GP acknowledges useful discussions with A. Franchi, L. Giacomelli, G. Santoro, and F. Tarantelli. MC acknowledges useful discussions with A. Solfanelli, G. Giachetti, N. Defenu and S. Ruffo on a problem that is closely related to the present work. 

This work has been partly supported by the Italian MIUR through PRIN Project No.~2017E44HRF.
Simulations have been performed using the Armadillo c++ library~\cite{Armadillo1, Armadillo2}.

\appendix

\section{The quantum Ising chain}
\label{App:ising}

In this appendix we briefly recall how to diagonalize the quantum Ising chain
of Eq.~\eref{Eq:ising}~\cite{Sachdev2011, Mbeng2020}.
First we introduce the Jordan-Wigner transformation
\begin{equation}
  \hat \sigma^+_j = \exp \bigg( i \pi \sum_{\ell = 1}^{j-1} \hat a^\dagger_{\ell}\hat a_{\ell} \bigg) \hat a_j,
  \label{eq:JW}
\end{equation}
with $\hat \sigma^\pm_j = \frac{1}{2}\big(\hat  \sigma_j^x \pm i \hat \sigma_j^y \big)$ denoting the
rising and lowering operators of the $j$th spin, and $\hat a_j^{(\dagger)}$ being anticommuting
fermionic annihilation (creation) operators, $\{ \hat a_i, \hat a^\dagger_j \} = \delta_{ij}$ and $\{ \hat a_i, \hat a_j\} = 0$.
This transformation maps Eq.~\eref{Eq:ising} into the spinless-fermion Hamiltonian
\begin{equation}
  \hat H = - J \! \sum_{j = 1}^{N-1} \!\! \left(\hat a^\dagger_j \hat a_{j+1}
  + \hat a^\dagger_j \hat a^\dagger_{j+1} + {\rm h.c.} \right)
  + h\sum_{j=1}^N \left( 2 \hat a^\dagger_j \hat a_j \!-\!1 \right) \!.
  \label{Eq:ising_fermions}
\end{equation}
By introducing the $2N$-dimensional Nambu spinor
$\mathbf{\hat a} =(\hat a_1, \dots,\hat a_N, \hat a^\dagger_1, \dots, a_N^\dagger)^T$,
such Hamiltonian can be written in the compact form
\begin{equation}
  \hat H = \frac{1}{2} \mathbf{\hat a}^\dagger \mathbb{H} \mathbf{\hat a} + {\rm const} \,,
\end{equation}
with the matrix 
\begin{subequations}
  \label{Eq:BdG}
  \begin{equation}
    \mathbb{H} = \left(\begin{array}{cc} A & B \\ B &A \end{array}\right)
  \end{equation}
  denoting the so-called Bogoliubov-de Gennes Hamiltonian with entries
  \begin{equation}
    \left \{ \begin{array}{cc} A_{j,j} =  h, & A_{j, j+1} = A_{j+1, j} \ \ = - J/2 \\		   
      B_{j,j} = 0, & B_{j, j+1} =- B_{j+1, j} = - J/2.\end{array} \right.
  \end{equation}
\end{subequations}

The above Hamiltonian can be diagonalized by defining a new $2N$-dimensional Nambu spinor
\begin{equation}
  \mathbf{\hat b} = (\hat b_1, \dots, \hat b_N, \hat b_1^\dagger, \dots, \hat b_N^\dagger) ,
  \label{eq:normal}
\end{equation}
where $\{\hat b_k^{(\dagger)}\}$ is another set of fermionic-quasiparticle operators,
through the relation $\mathbf{\hat b} = \mathbb{U}^{-1} \mathbf{ \hat a}$. 
The matrix
\begin{equation}
  \mathbb{U} = \left(\begin{array}{cc} U & V^* \\ V & U^* \end{array}\right)
  \label{Eq:bogo}
\end{equation}
expresses a so-called Bogoliubov transformation, and is such that
\begin{equation}
  \label{eq:bogodiag}
  \mb{U}^{-1} \, \mb{H} \, \mb{U}  =  {\rm diag} \big( \omega_k, -\omega_k \big).
\end{equation}
By imposing that the $\hat b_k$ fermions satisfy fermionic commutation relations,
one obtains the following constraints for the blocks of the Bogoliubov transformation~\eref{Eq:bogo}:
\begin{equation}
  U U^\dagger + V V^\dagger = \mb{I}, \quad UV^T + VU^T = 0.
  \label{Eq:commutations}
\end{equation}
In the thermodynamic limit, the dispersion relation entering Eq.~\eref{eq:bogodiag}
has the analytic expression
\begin{equation}
  \label{eq:dispersion}
  \omega_k  =  2J \sqrt{1 + \left( \frac{h}{J} \right)^2
    - 2 \left( \frac{h}{J} \right) \cos \left(k \right) },
\end{equation}
where $k \in [0, 2 \pi)$ is a real number denoting the fermionic quasimomentum.
The same expression holds for finite-size systems with periodic boundary conditions,
by considering $k \in \mathcal{K}$, being $\mathcal{K}$ a parity-sector-depending discrete set~\cite{Mbeng2020}.
In contrast, it is not possible to find an analytic expression in the case with open boundary conditions.

Equation~\eref{eq:dispersion} gives the energies of the $\hat b_k$ fermionic
quasiparticles ($\omega_k \geq 0$), therefore it is clear that the ground state $\ket{\psi}_{\rm g.s.}$
of the system is the vacuum $\ket{0}$ of such fermions.
Transforming back to the original $\hat a_j$ fermions of Hamiltonian~\eref{Eq:ising_fermions},
one can write
\begin{equation}
  \ket{\psi}_{\rm g.s.} =
  \mathcal{N} \exp \bigg( \frac{1}{2} \sum_{j, l=1}^N Z_{jl} \hat a_j \hat a_l \bigg) \ket{0},
  \label{Eq:gauss}
\end{equation}
where $Z = - (U^\dagger)^{-1} V^\dagger$ and $\mathcal{N}$ is a normalization factor. 
The associated ground-state energy
\begin{equation}
  E_{\rm g.s.} = - \sum_{k \in \mathcal{K}} \omega_k
\end{equation}
is a decreasing function of the transverse field $h$.
Likewise all the excitation spectrum can be easily obtained by progressively populating
the vacuum $\ket{0}$ with the quasiparticles raising operators $\hat b^\dagger_k$,
each of them associated with an energy $\omega_k$. 

Because of the Gaussian form of Eq.~\eref{Eq:gauss}, the state is fully determined by the two-point
correlation functions $G_{jl} = \braket{\hat a_j \hat a^\dagger_l}$ and $F_{jl} = \braket{\hat a_j \hat a_l}$,
defined through
\begin{equation}
  \mathbb{G} = \mathbb{U} \left(\begin{array}{c|c}  \mathbb{I} & 0 \\ \hline 0 & 0 \end{array}\right) \mathbb{U}^\dagger = \left(\begin{array}{c|c}  G & F \\ \hline F^\dagger & 1-G^T \end{array}\right).
  \label{Eq:corr}
\end{equation}
Since the Hamiltonian~\eref{Eq:ising_fermions} is quadratic in the fermionic operators,
the knowledge of $\mathbb{G}$ directly gives access to the zero temperature 
Hamiltonian expectation values.
Moreover, the application of any operator that is a quadratic function of $\hat a^{(\dagger)}$ leaves
the Gaussian form invariant. 
Therefore, if one wants to study the unitary dynamics starting from $\ket{\psi}_{\rm g.s.}$
and following a variation of the Hamiltonian parameters $J \equiv J(t)$ and $h \equiv h(t)$,
this can be done by just tracking the evolution of the matrix $\mb{U}$:
\begin{equation}
  i \, \partial_t \mb{U}(t) = 2 \, \mb{H}(t) \, \mb{U}(t), 
  \label{Eq:unitary}
\end{equation}
being $\mb{H}$ the Bogoliubov-de Gennes Hamiltonian defined in Eqs.~\eref{Eq:BdG}. 
By substituting the solution of Eq.~\eref{Eq:unitary} in Eq~\eref{Eq:corr}, it is possible
to evaluate the correlations at time $t$ and, consequently, the expectation value
$\braket{\hat H(t)}_{\rho(t)}$ of the Hamiltonian at time $t$, with $\rho(t) = \ket{\psi(t)}\!\bra{\psi(t)}$.

Before concluding we mention that, even though the discussion above is for pure states, because of
the Gaussianity of the model, the same formalism can be adopted to thermal states as well.
In fact thermal states are simply mixtures of pure states,
weighted by the corresponding Boltzmann factor, and thus keep a Gaussian character.
As a consequence [see the discussion in~\ref{sec:therm} and, in particular, Eq.~\eref{eq:corr_T}],
averages over thermal Gaussian states are obtained by combining two contributions, one that accounts
for ground-state correlations (i.e., the matrix $\mathbb{G}$) and the other that accounts for thermal effects.

\section{Thermalisation stroke}
\label{sec:therm}

In this section we present some details of the nonlocal master equation chosen to model
the system-environment interaction~\cite{Petruccione2007, DAbbruzzo2021}. 
Let us consider, for a while, the more general configuration of $N_b$ independent thermal reservoirs
at temperature $T_n$, with $n \in \{1, \dots, N_B\}$ indices labeling the bath.

The Hamiltonian describing this setup reads
\begin{equation}
  \hat H_{\rm env} = \sum_{n = 1}^{N_B} \int dk \, \epsilon_n(k) \, \hat c_n^\dagger(k) \, \hat c_n(k),
\end{equation}
with $\epsilon_n(k) \ge 0$ and $\hat c_n^{(\dagger)}$ fermionic annihilation (creation) operators.
The $N_b$ baths are independent, therefore the corresponding reduced density operator
of the full environment assumes the factorized form 
\begin{equation}
  \rho_{\rm env} = \bigotimes_{n=1}^N \rho_{\rm bath}^{(n)}, 
\end{equation}
being $\rho_{\rm bath}^{(n)}$ the thermal density matrix describing the $n$th fermionic bath
at temperature $T_n$.

Let us assume the $n$th of these baths to be coupled to $p$ system sites,
and define $\mathcal{I}_n$ as the ensemble of these points. 
The coupling between the system and the environment is described by a quadratic factorizable Hamiltonian
\begin{equation}
  \hat H_{\rm int} \! = \sum_{n=1}^{N_B} \sum_{p\in \mathcal{I}_n} \! \int \! dk \ g_n(k)
  \big( \hat a_p + \hat a_p^\dagger \big) \, \big[ \hat c_n(k) + \hat c_n^\dagger(k) \big] ,
\end{equation}
where $\{ \hat a_p^{(\dagger)} \}$ are the fermionic operators of the system, as defined in Eq.~\eref{eq:JW},
while $g_n(k)$ quantifies the interaction strength between the $k$th mode of the $n$th bath and the
sites $p\in \mathcal{I}_n$ of the system.
This equation can be written in a factorized form $\hat H_{\rm int} = \sum_{n= 1}^{N_B} \hat O_n \otimes \hat R_n$ by posing
$\hat O_n = \sum_{p \in \mathcal{I}_n} \big( \hat a_p + \hat a_p^\dagger \big)$ and
$\hat R_n =\int dk \ g_n(k) \big[ \hat c_n(k) + \hat c_n^\dagger(k) \big]$.
In what follows is useful to introduce the density of states associated with the $n$th bath:
\begin{equation}
  \mathcal{J}_n(\omega) \equiv \pi \int dk \ |g_n(k)|^2 \,\, \delta \big[\omega - \epsilon_n(k) \big] .
\end{equation}
Under the assumption that the baths have a very large bandwidth with respect to the frequencies
of the system, we have that $\mathcal{J}_n(\omega) \simeq \mathcal{J}_n$.

Tracing out all the environmental degrees of freedom and imposing the Bork-Markov approximation for the baths, it is possible to derive a microscopic
Lindblad master equation~\cite{Petruccione2007}, in the energy eigenbasis, for the reduced density matrix of the system
described by the Hamiltonian~\eref{Eq:ising_fermions}~\cite{DAbbruzzo2021}:
\begin{subequations}
  \label{Eq:Lind}  
  \begin{equation}
    \partial_t {\rho}_{\rm sys}(t) = -i [\hat H_{\rm sys}, \rho_{\rm sys}] + \mathcal{D}[\rho_{\rm sys}],
  \end{equation}
  with
  \begin{eqnarray}
    D[\rho_{\rm sys}] &= & \sum_{n,k} \gamma_{nk} \! \left[ (1 \! - \! f_n(\omega_k) ) \left( 2\hat b_k \rho_{\rm sys} \hat b^\dagger_k \! - \! \{ \hat b^\dagger_k \hat b_k, \rho_{\rm sys} \} \right) \right] \nonumber\\ 
    &	+ &\sum_{n,k} \gamma_{nk}\left[f_n(\omega_k) \! \left(2 \hat b_k^\dagger \rho_{\rm sys} \hat b_k - \{ \hat b_k \hat b_k^\dagger, \rho_{\rm sys} \} \right) \right],
  \end{eqnarray}
\end{subequations}
where the $\{ b_k^{(\dagger)} \}$ jump operators are the fermionic Bogoliubov quasiparticles
which diagonalize the model in Eq.~\eref{Eq:ising_fermions}.
Of course, these operators are local in the energy eigenbasis and thus nonlocal in the sites,
giving rise to a global Master equation. 
Moreover
\begin{subequations}
  \begin{eqnarray}
    f_n(\omega_k) & = & \big(1 + e^{\omega_k/T_n} \big)^{-1} \,, \\
    \gamma_{nk} & = & \mathcal{J}_n \sum_{p, s \in \mathcal{I}_n} (U_{pk} + V_{p k})(U_{ks}^* + V_{ks}^*)\,, \qquad
  \end{eqnarray}
\end{subequations}
are, respectively, the Fermi-Dirac distribution function and the bath coupling constants,
$U, \, V$ being the Bogoliubov matrices of Eq.~\eref{Eq:bogo}.
We observe that Eq.~\eref{Eq:Lind} maintains a quadratic structure in the fermionic operators
so that it can be handled with Bogoliubov techniques and thus efficiently simulated
for systems with hundreds of sites~\cite{DAbbruzzo2021}.

Under the assumption of no degeneracies in the spectrum (as turns out to be the case, for the
Ising chain with open boundary conditions), Eq.~\eref{Eq:Lind} can be used to obtain an analytic
expression for the time evolution of the correlation functions.
In particular, defining $\tilde{f}_k = \frac{\sum_n \gamma_{nk} f_n(\omega_k)}{\sum_n \gamma_{nk}}$, we have
\begin{equation}
  \braket{\hat b_k^\dagger \hat b_k}_t = \tilde{f}_k \big(1 - e^{-2\sum_n \gamma_{nk}t} \big)
  + \braket{\hat b_k^\dagger \hat b_k}_0 e^{-2\sum_n \gamma_{nk}t}.
  \label{eq:corr_t}
\end{equation}
This expression suggests the existence of a mode-dependent thermalisation time
\begin{equation}
  t^\star_k \sim \Big( \sum_n \gamma_{nk} \Big)^{-1},
\end{equation}
after which the system reaches a unique thermal stationary state,
$\braket{\hat b^\dagger_k \hat b_k}_{\infty} = \tilde{f}_k$.
In general, there is no reason to expect a finite thermalisation time, namely $\sum_n \gamma_{nk} \ne 0$. 
However, assuming to have $N_b = N$ baths at the same temperature $T$, with the same density of state
$\mathcal{J}_n = \mathcal{J}$, each of them coupled only to one spin ($\mathcal{I}_n = \{ n\}$),
we obtain a mode-independent thermalisation time $t^\star \sim 1/\mathcal{J}$. 
This simply comes from the relations in~\eref{Eq:commutations} and the fact that
\begin{eqnarray}
  \sum_n \gamma_{nk} & = & \mathcal{J}\sum_n \left(U_{nk} + V_{nk} \right)\left(U_{nk}^* + V_{nk} \right) \nonumber \\
  & = & \mathcal{J} \Big[ \! \left(UU^\dagger \! + \! VV^\dagger\right)_{kk}
    + \left(UV^T \! + \! VU^T\right)_{kk} \Big] = \mathcal{J}. 
\end{eqnarray}
Since all the baths are identical (independent of the mode $k$), the stationary state
is precisely the thermal state at temperature $T$. 

Once thermalisation is reached, finite-temperature correlation functions
(and, consequently, Hamiltonian expectation values) can be evaluated by virtue
of the following relations 
\begin{eqnarray}
  \braket{\hat a^\dagger_i \hat a_j}_T &
  = & \! (U^*(t) \, \Theta \, U^T(t) - V(t) \, \Theta \, V^\dagger(t) + V(t) \, V^\dagger(t))_{ij}, \nonumber\\
  \braket{\hat a_i^\dagger \hat a_j^\dagger}_T &
  = & \! (U^*(t) \, \Theta \, V^T(t) - V(t) \, \Theta \, U^\dagger(t) + V(t) \, U^\dagger(t))_{ij}, 
  \label{eq:corr_T}
\end{eqnarray}
where $\braket{\, \cdot \,}_T$ indicates the average over the thermal state at temperature $T$ and $\Theta_{kq} = \delta_{kq}f_k$.

For the bath configuration we chose in our model, we are ensured that the system eventually thermalises. 
For this reason, to derive the results in the main text, we needed Eq.~\eref{eq:corr_T} only. 
However, the above formalism would also apply to more generic non-thermalising situations,
allowing to characterise the relaxation process of the system coupled to the environment,
through the time evolution of the Hamiltonian expectation values.
In fact, the correlation functions at a generic time $t$ (after the system has been put in contact with the reservoir)
are obtained by substituting $\Theta_{kq} \mapsto \Theta_{kq}^t = \delta_{kq}\braket{\hat b^\dagger_k \hat b_k}_t$
in Eq.~\eref{eq:corr_T}, where $\braket{\hat b^\dagger_k \hat b_k}_t$ are the correlators
defined in Eq.~\eref{eq:corr_t}.
%

\section*{References} %

\bibliography{critical_otto}

\end{document}